\RequirePackage{fix-cm}
\documentclass[twocolumn]{svjour3}          
\smartqed  
\usepackage{graphicx}
\usepackage{indentfirst}
\usepackage{amsmath}
\usepackage{amsfonts}
\begin{document}

\title{Generalized model for steady-state bifurcations without parameters in memristor-based oscillators with lines of equilibria
}


\author{lvan~A.~Korneev         \and
	    Andrei~V.~Slepnev      \and
	    Anna~S.~Zakharova    \and
        Tatiana E. Vadivasova \and
        Vladimir V. Semenov 
}


\institute{
I.A. Korneev, A.V. Slepnev, T.E. Vadivasova \at 
Saratov State University, Astrakhanskaya str., 83, 410012, Saratov, Russia \\
           	\and
A.S. Zakharova \at 
Technische Universit\"{a}t Berlin, Hardenbergstra{\ss}e 36, 10623, Berlin, Germany \\
		\and
V.V. Semenov \at 
Technische Universit\"{a}t Berlin, Hardenbergstra{\ss}e 36, 10623, Berlin, Germany \\
Saratov State University, Astrakhanskaya str., 83, 410012, Saratov, Russia \\
              \email{semenov.v.v.ssu@gmail.com}    
}

\date{Received: date / Accepted: date}

\maketitle

\begin{abstract}
We demonstrate how the pitchfork, transcritical and saddle-node bifurcations of steady states observed in dynamical systems with a finite number of isolated equilibrium points occur in systems with lines of equilibria. The exploration is carried out by using the numerical simulation and linear stability analysis applied to a model of a memristor-based oscillator. First, all the discussed bifurcation scenarios are considered in the context of systems including Chua's memristor with a piecewise-smooth characteristic. Then the memristor characteristic is changed to a function that is smooth everywhere. Finally, the action of the memristor forgetting effect is taken into consideration. The presented results are obtained for electronic circuit models, but the considered bifurcation phenomena can be exhibited by  systems with a line of equilibria of any nature.
\end{abstract}

\section{Introduction}
\label{intro}
The simultaneous and continuous dependence of the oscillatory dynamics both on parameter values and initial conditions is a frequent occurrence in various dynamical systems. In most cases, this peculiarity is associated with the properties of conservative oscillators. However, such behaviour can be exhibited by nonlinear dissipative oscillators with manifolds of equilibria. The manifolds of equilibria consist of non-isolated equilibrium points and takes different forms: line of equilibria \cite{fiedler2000-1,fiedler2000-2,fiedler2000-3,liebscher2015}, surface of equilibria \cite{jafari2016}, circle \cite{gotthans2015,gotthans2016} and square \cite{gotthans2016} equilibria, etc. Among such manifolds one distinguishes $m$-dimensional normally hyperbolic manifolds of equilibria characterized by $m$ pure imaginary eigenvalues or $m$ eigenvalues being equal to zero, whereas all the other eigenvalues have non-zero real parts. In the simplest case these manifolds exist as a line of equilibria. Systems with a line of equilibria have been mathematically considered \cite{fiedler2000-1,fiedler2000-2,fiedler2000-3,liebscher2015,riaza2012,riaza2016,corinto2020}. It has been shown that their significant feature is the occurrence of so-called bifurcations without parameters, i.e., the bifurcations corresponding to fixed parameters when the condition of normal hyperbolicity is violated at some points of the line of equilibria. 

Memristor-based oscillators are widely represented in a variety of systems with a line of equilibria. One of the frequently considered models is the series RLC-circuit with negative resistance, where a flux-controlled memristor is connected in parallel with the capacitor. Bifurcation mechanisms of the periodic solution appearance in this system have been explored numerically and analytically for different kinds of nonlinearity \cite{messias2010,botta2011,riaza2012,semenov2015,korneev2017,korneev2017-2,korneev2021-2}. Analytical solution of the model equations has enabled to recognize significant features of the supercritical \cite{korneev2017,korneev2017-2} and subcritical \cite{korneev2021-2} Andronov-Hopf bifurcation. In addition, the recently published research \cite{korneev2021-2} has demonstrated how the saddle-node bifurcation of limit cycles observed in classical self-oscillators with hard self-oscillation excitation transforms in systems with a line of equilibria. Thus, all the bifurcations  of limit cycles exhibited by self-oscillators with one degree of freedom have been described in terms of oscillators with a line of equilibria. The similarity between classical self-oscillators and the considered systems with a line of equilibria is emphasized by the fact that undamped periodic oscillations in systems with a line of equilibria can be synchronized by external periodic forcing \cite{korneev2020}. 

In contrast to our previous publications focused on the self-oscillatory dynamics, in the current paper we show how classical bifurcations of steady states are manifested in systems with a line of equilibria. Three bifurcations are studied: the pitchfork bifurcation, the transcritical bifurcation and the saddle-node bifurcation of steady states. It is demonstrated that the bifurcations under study become bifurcations without parameters due to the memristor properties. Our purpose is to describe all the models and bifurcations in a unified manner. For this reason, all the considered systems belong to the same class of circuits, the RLC-circuit with feedback including the memristor. 

\section{Model and methods}
\label{model}

\begin{figure*}[t]
\centering
\includegraphics[width=0.8\textwidth]{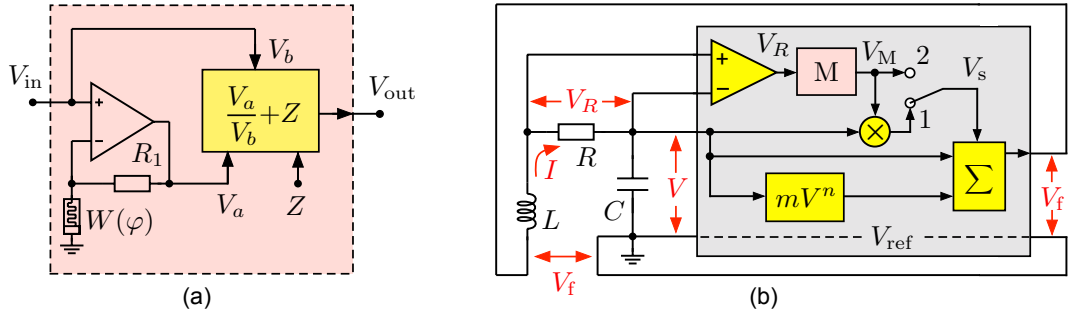}
\caption{(a) Memristive circuit based on the memristive non-inverting amplifier and analog divider; (b) Schematic circuit diagram of the studied model (Eqs.(\ref{physical_system})).}
\label{fig1}
\end{figure*}

According to the idea proposed by Leon Chua \cite{chua1971} the memristor relates the transferred electrical charge, $q(t)$, and the magnetic flux linkage, $\varphi(t)$, by means of the linear relationship $dq=Wd\varphi$, whence it follows that $W=W(\varphi)=\dfrac{dq}{d\varphi}$. By this way, using the relationships $d\varphi=V_{\text{m}}dt$ and $dq=I_{\text{m}}dt$ ($V_{\text{m}}$ is the voltage across the memristor, $I_{\text{m}}$ is the current passing through the memristor), the memristor current-voltage characteristic can be derived: $I_{\text{m}}=W(\varphi)V_{\text{m}}$. In such a case, $W$ is the flux-controlled conductance (memductance) and depends on the entire past history of $V_{\text{m}}(t)$: 
\begin{equation}
W(\varphi)=\dfrac{dq}{d\varphi}=q '\left( \int\limits_{-\infty}^{t}{V_{\text{m}}(t)dt} \right).
\label{W(phi)}
\end{equation}
It must be noted that the term 'flux' (or 'flux linkage') is used below only for denoting the memristor state variable being proportional to the integral $\int\limits_{-\infty}^{t}{V_{\text{m}}(t)dt}$. The issues concerning the physical realizability of the postulated relationship $dq=Wd\varphi$ are not discussed in the current paper. Thus, the memristor is considered as a resistive component which conductance is dictated by the state variable which is not necessarily associated with magnetic phenomena. This approach reflects the conception of 'memristive system' \cite{chua1976} implying the mathematical definition, which does not concern a physical sense of the dynamical variables and their functional dependence. It allows for grouping a broad variety of elements of different nature identified by a continuous functional dependence of characteristics on previous states. 

One of the simplest memristor model is Chua's memristor described by the piecewise-linear dependence $q(\varphi)$ which takes the following form for the flux-controlled memristor:
\begin{equation}
q(\varphi)=
\begin{cases}
	  (a-b)\varphi_{*}+b\varphi, & \varphi \ge \varphi_{*},\\
          a \varphi , & |\varphi| < \varphi_{*},\\
          -(a-b)\varphi_{*}+b\varphi , & \varphi \le -\varphi_{*}.
\end{cases}
\label{q_phi_chua_memristor}
\end{equation}
Then the memristor conductance $W(\varphi)$ becomes:
\begin{equation}
W(\varphi)=
\begin{cases}
          a , & |\varphi| < \varphi_{*},\\
          b , & |\varphi| \geq \varphi_{*}.
\end{cases}
\label{chua_memristor}
\end{equation}
Nonlinearity (\ref{chua_memristor}) can be approximated by the hyperbolic tangent function:
\begin{equation}
W(\varphi)=\dfrac{b-a}{2} \tanh\left(k(\varphi^2-\varphi_*)\right)+\dfrac{b+a}{2},
\label{tanh_memristor}
\end{equation}
where a parameter $k$ characterizes the sharpness of the transitions between two memristor's states. It has been shown in \cite{korneev2021-2} that changing the memristor conductance function to the smooth one does not qualitatively modify the memristor properties. The classical loop in the current-voltage characteristic of the memristor driven by the external periodic influence persists (see Fig. 2 in Ref. \cite{korneev2021-2}). The memristor model including tanh-nonlinearity is not the only smooth memristor model. There is a number of smooth models describing various memristor properties \cite{tetzlaff2014,linn2014,singh2019,ascoli2013-2,chang2011,chua2011,guseinov2021}.

Real memristive systems can 'forget' the state history over time. In particular, the 'forgetting' effect in memristors based on metal oxides is associated with the diffusion of charged particles \cite{chang2011,chen2013,zhou2019} (however, the 'forgetting' can happen very slowly). One of the simplest form of the memristor state equation which implies the forgetting effect is the following:
\begin{equation}
\label{memristor_with_forgetting}
\dfrac{dz}{dt}=g(x,z)=x-\delta z,
\end{equation}
where $z$ plays a role of a memristor state variable, $x$ is an input signal, a parameter $\delta$ characterizes the forgetting effect strength.

Consider the circuit in Fig.~\ref{fig1}~(a). It contains the operational-amplifier-based non-inverting memristive amplifier which output signal is $V_{a}=(1+R_{1}W(\varphi))V_{\text{in}}$, where $\varphi$ varies according to memristor state equation (\ref{memristor_with_forgetting}) taken in the form $\dfrac{d\varphi}{dt}=g(V_{\text{in}},\varphi)$. The second block is an analog divider producing the output voltage such that the resulting circuit response is $V_{\text{out}}=1+Z+R_{1}W(\varphi)$, where $Z$ is a summing input voltage. The memristive circuit in Fig.~\ref{fig1}~(a) is included as the memristive block $M$ into the schematic circuit diagram in Fig.~\ref{fig1}~(b). The circuit in Fig.~\ref{fig1}~(b) represents the linear series RLC-circuit forced by the feedback signal $V_{\text{f}}(t)$ being an output signal of the complex nonlinear amplifier (the grey block in Fig.~\ref{fig1}~(b)). The amplifier has two inputs: the voltage drop across the resistor, $V_R=RI$ and the voltage across the capacitor, $V$. The amplifier input currents are assumed to be zero. After the voltage $V_R$ is reproduced by a differential amplifier, it becomes an input signal of the memristive circuit $M$ which output $V_{\text{M}}$ is comes to the analog multiplier and to the two-state switch. If the switch state is 1, the corresponding output signal is $V_{\text{s}}=V_{\text{M}}V$. Otherwise, the signal coming to the summing block $\sum$ is $V_{\text{s}}=V_{\text{M}}$. In addition, the amplifier contains the nonlinear block for the square or cubic transformation $sV^n$ of the input. Thus, the amplifier output signal is $V_{\text{f}}=V_{\text{s}}+V+mV^n$. The presented in Fig.~\ref{fig1}~(b) system is described by the following dynamical variables: $V$ is the voltage across the capacitor $C$,  $I$ is the current through the inductor $L$ and $\varphi$ is the magnetic flux linkage controlling the memristor. Using Kirchhoff’s laws, one obtains differential equations for the considered system in physical time $t'$:
\begin{equation}
\label{physical_system}
\left\lbrace
\begin{array}{l}
C\dfrac{dV}{dt'}=I,\\
\\
L\dfrac{dI}{dt'}=-RI+V_{\text{s}}(V,V_R,\varphi)+mV^n,\\
\\
\dfrac{d\varphi}{dt'}=RI - k \varphi,
\end{array}
\right.
\end{equation}
where the third equation reflects the memristor forgetting effect. In the dimensionless variables $x=V / V_{0}$, $Y=I/ I_{0}$ and $z=\varphi/(L\varphi_{0})$ with $V_{0}= 1$~V, $I_{0}=1$~A, $\varphi_{0}=\text{1 sec} \times V_{0}$ and the dimensionless time $t=[(V_{0}/(I_{0}L)]t'$, Eqs.(\ref{physical_system}) can be rewritten in the following form:
\begin{equation}
\left\lbrace
\begin{array}{l}
\dfrac{dx}{dt}=\nu Y,\\
\\
\dfrac{dY}{dt}=-\gamma Y+f(x,Y,z)+\mu x^n,\\
\\
\dfrac{dz}{dt} = \gamma Y-\delta z, \\
\end{array}
\right.
\label{system}
\end{equation}
where $\nu=(L/C)(I_{0}/V_{0})^{2}$  is a dimensionless parameter being numerically equal to $L/C$, $\gamma=R(I_{0}/V_{0})$ is a dissipation factor, $f(x,Y,z)$ is a function reflecting the memristive properties of the block $M$ in Fig.~\ref{fig1}~(d) and represents a dimensionless analog of the signal $V_{\text{s}}$, $\mu$ is a dimensionless equivalent of the parameter $m$, $\delta=kL(I_{0}/V_{0})$. The substitution $y=\gamma Y$ transforms model (\ref{system}) into the finalized form:
\begin{equation}
\left\lbrace
\begin{array}{l}
\dfrac{dx}{dt}=\alpha y,\\
\\
\dfrac{1}{\gamma}\dfrac{dy}{dt}=-y+f(x,y,z)+\mu x^n,\\
\\
\dfrac{dz}{dt} = y-\delta z, \\
\end{array}
\right.
\label{system_final}
\end{equation}
where $\alpha=\nu/\gamma$. The function $f(x,y,z)$ contains the functional dependence on the memristor instantaneous state. In physical variables the dependence is described as the expression for the memristor conductance given either by  Exps. (\ref{chua_memristor}) or (\ref{tanh_memristor}). In the context of model (\ref{system_final}),  the function $W_{M}(\varphi)$ corresponds to the equivalent dimensionless form $G_M(z)$. Two options for $G_M(z)$ are under consideration:
\begin{equation}
G_M(z)=
\begin{cases}
          a , & |z| < z_0,\\
          b , & |z| \geq z_0,
\end{cases}
\label{chua_memristor_dimensionless}
\end{equation}
and
\begin{equation}
G_M(z)=\dfrac{b-a}{2} \tanh\left(k(z^2-z_0)\right)+\dfrac{b+a}{2}.
\label{tanh_memristor_dimensionless}
\end{equation}

System (\ref{system_final}) is explored both theoretically by using the linear stability analysis and numerically by means of integration methods. Numerical simulations are carried out by integration of Eqs. (\ref{system_final}) using the fourth-order Runge-Kutta method with the time step  $\Delta t = 0.0001$ from different initial conditions. Particular modifications of dynamical model (\ref{system_final}) are considered below in more details.

\section{Pitchfork bifurcation}
Suppose that the switch in Fig.~\ref{fig1}~(b) operates in state 1 while the nonlinear transformation is cubic. Then the expression for the feedback voltage takes the form $V_{\text{f}}=(1+Z+R_1W(\varphi))V+V+mV^3$, where the state variable varies according the equation $d\varphi /dt =RI-k\varphi$ and $m=-1$. Adjusting the voltage $Z$ and the resistance $R_1$, one controls the coefficient $(1+Z+R_1W(\varphi))$ which can possess both negative and positive values. Then the physical model (see Eqs.~(\ref{physical_system})) corresponding to the presence of piecewise-smooth memristor (\ref{chua_memristor_dimensionless}) or its continuous analog (\ref{tanh_memristor_dimensionless}) is transformed to the dimensionless form 
\begin{equation}
\left\lbrace
\begin{array}{l}
\dfrac{dx}{dt}=\alpha y,\\
\\
\dfrac{1}{\gamma}\dfrac{dy}{dt}=-y+\beta(z)x-x^3,\\
\\
\dfrac{dz}{dt} = y-\delta z, \\
\end{array}
\right.
\label{pitchfork_model}
\end{equation}
for two options $\beta(z)$: 
\begin{equation}
\beta(z)=
\begin{cases}
          \beta_1 , & |z| < 1,\\
          \beta_2 , & |z| \geq 1,
\end{cases}
\label{beta_piecewise}
\end{equation}
\begin{equation}
\beta(z)=\dfrac{\beta_2-\beta_1}{2} \tanh\left(k(z^2-1)\right)+\dfrac{\beta_2+\beta_1}{2}.
\label{beta_tanh}
\end{equation}
System (\ref{pitchfork_model}) is studied for fixed parameters: $\alpha=\gamma=1$, $\beta_1=-1$, $\beta_2=1$, $k=5$. First, the memristor forgetting effect is excluded from the consideration: $\delta=0$. 

\begin{figure}
\centering
\includegraphics[width=0.5\textwidth]{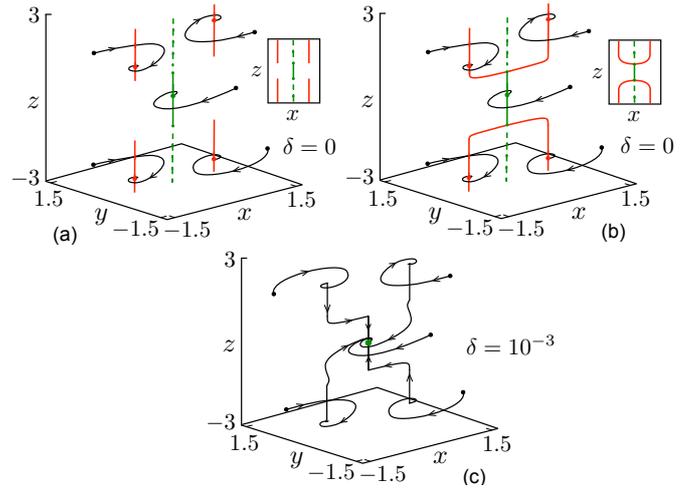} 
\caption{Pitchfork bifurcation without parameters in the phase space of system (\ref{pitchfork_model}) for memristive nonlinearities (\ref{beta_piecewise}) (panel (a)) and (\ref{beta_tanh}) (panel (b)) and the absence of the forgetting effect, $\delta=0$. The attracting lines of equilibria are marked by solid green and red lines, the unstable branches of the central line of equilibria are marked by green dashed lines. Phase trajectories are black arrowed curves. The right insets schematically show the manifolds of equilibria as a projection on the plane $(x,z)$; (c) Phase trajectories in system  (\ref{pitchfork_model})-(\ref{beta_tanh}) for $\delta=10^{-3}$ tracing motions to the stable steady state (green point) in the origin. Other system parameters are: $\alpha=\gamma=1$, $\beta_1=-1$,$\beta_2=1$, $k=5$.}
\label{fig2}
\end{figure}  

System (\ref{pitchfork_model}) with piecewise-smooth function (\ref{beta_piecewise}) does not contain the variable $z$ in an explicit form. Then two cases $\beta=\beta_1$ and $\beta=\beta_2$ are considered independently. In case $|z| < 1$ the system attractor is a manifold of steady states with the coordinates $x_1=0$, $y_1=0$, $z\in(-1;1)$ (the green solid line in Fig.~\ref{fig2}~(a)). All the trajectories in the subspace $z\in(-1;1)$ are attracted to the manifold of equilibria. Thus, the attractor represents a line of equilibria characterised by the eigenvalues $\lambda_1=0$, $\lambda_{2,3}=(-1\pm i\sqrt{3})/2$ considered as stable fixed points since $\lambda_{2,3}$ have negative real part. Each point of the line of equilibria is neutrally stable in the OZ-axis direction. Hereinafter, using the terms 'stable' or 'unstable' point at the line of equilibria, we mean the behaviour of trajectories (attraction or repelling) in the neighbourhood of the equilibrium point, which is determined by the eigenvalues $\lambda_{2,3}$. 

For two subspaces $|z| \geq 1$, there exist three lines of equilibria. The first one ($x_1=0$, $y_1=0$, $|z|\geq 1$) represent a line of saddle-like equilibria (the green dashed line in Fig.~\ref{fig2}~(a)) with eigenvalues $\lambda_1=0$, $\lambda_{2,3}=(-1\pm \sqrt{5})/2$. The second and third lines of equilibria (the red solid lines in Fig.~\ref{fig2}~(a)) are two coexisting attractors which consist of steady states ($x_{2,3}=\pm1$, $y_{2,3}=0$, $|z| \geq 1$) with the corresponding eigenvalues $\lambda_1=0$, $\lambda_{2,3}=(-1\pm i\sqrt{7})/2$. Thus, for $|z| \geq 1$ the central line of equilibria plays a role of an unstable fixed points   between two attractors in classical bistable oscillators. 

In summary, changing $z$ and keeping parameter values to be fixed, one implements the transition between the existence of a single attractor to the regime of bistablility which consists in the coexistence of two stable lines of equilibria. The described bifurcation transition corresponds to the pitchfork bifurcation which occurs at $z=\pm1$: the line of equilibria becomes unstable for $|z|\geq 1$ and two coexisting attractors appear at the moment of bifurcation. If system (\ref{pitchfork_model}) involves tanh-function (\ref{beta_tanh}), the similarity with the classical pitchfork bifurcation is more evident [Fig.~\ref{fig2}~(b)]: there exists a single line of equilibria in the central subspace, which loses stability when $z$ passes through the threshold value $z_{*}\approx \pm1$. At the bifurcation moment two stable lines of equilibria begin in the vicinity of the unstable line of equilibria. 

For non-zero values of the parameter $\delta$ the lines of equilibria disappear [Fig.~\ref{fig2}~(c)]. In such a case, the steady state $x=y=z=0$ becomes a single attractor in the phase space. Then the transition to the bistability occurs in a classical way by varying parameter $\beta_1$: the pitchfork bifurcation occurs at $\beta_1=0$. However, for small values $\delta$ the phase trajectories describing motion to the stable fixed point can trace the lines of equilibria initially existed at $\delta=0$ (see the trajectories in Fig.~\ref{fig2}~(c)). 

\section{Transcritical bifurcation}
Suppose that the switch in Fig.~\ref{fig1}~(b) operates in state 1 while the nonlinear transformation is quadratic. Then the expression for the feedback voltage becomes $V_{\text{f}}=(1+Z+R_1W(\varphi))V+V+mV^2$, where the state variable varies according the equation $d\varphi /dt =RI-k\varphi$ and $m=-1$. Following the same procedures as in the previous section, one obtains the dimensionless model of the circuit:
\begin{equation}
\left\lbrace
\begin{array}{l}
\dfrac{dx}{dt}=\alpha y,\\
\\
\dfrac{1}{\gamma}\dfrac{dy}{dt}=-y+\beta(z)x-x^2,\\
\\
\dfrac{dz}{dt} = y-\delta z, \\
\end{array}
\right.
\label{transcritical_model}
\end{equation}
where the function $\beta(z)$ has two options: Exp. (\ref{beta_piecewise}) or Exp. (\ref{beta_tanh}). System (\ref{transcritical_model}) is studied for fixed parameters: $\alpha=\gamma=1$, $\beta_1=-1$, $\beta_2=1$, $k=5$. 

\begin{figure}
\centering
\includegraphics[width=0.5\textwidth]{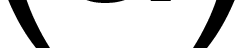} 
\caption{Transcritical bifurcation without parameters in the phase space of system (\ref{transcritical_model}) for memristive nonlinearities (\ref{beta_piecewise}) (panel (a)) and (\ref{beta_tanh}) (panel (b)) and the absence of the forgetting effect, $\delta=0$. The attracting segments of the lines of equilibria are marked by solid red lines while the repelling ones are marked by green dashed lines. Phase trajectories are black arrowed curves. The right insets schematically show the manifolds of equilibria as a projection on the plane $(x,z)$; (c) Phase trajectories in system  (\ref{transcritical_model})-(\ref{beta_tanh}) for $\delta=10^{-3}$ tracing motions to the stable steady state (green point). Other system parameters are: $\alpha=\gamma=1$, $\beta_1=-1$,$\beta_2=1$, $k=5$.}
\label{fig3}
\end{figure}

First, system (\ref{transcritical_model}) is considered for piecewise-smooth function $\beta(z)$ (function (\ref{beta_piecewise})) and $\delta=0$. Two cases $\beta=\beta_1$ and $\beta=\beta_2$ are considered independently. In case $|z| < 1$ the system attractor is a manifold of steady states with the coordinates $x_1=0$, $y_1=0$, $z\in(-1;1)$ characterised by the eigenvalues $\lambda_1=0$, $\lambda_{2,3}=(-1\pm i\sqrt{3})/2$ (the red solid line in Fig.~\ref{fig3}~(a)). All the phase trajectories in the subspace $z\in(-1;1)$ are attracted to the line of equilibria $x_1=0$, $y_1=0$, $z\in(-1;1)$. In addition, there exists the second line of equilibria ($x_2=-1$, $y_2=0$, $z\in(-1;1)$) (the green dashed line in Fig.~\ref{fig3}~(a)) exhibiting the properties of a saddle fixed point (the corresponding eigenvalues are $\lambda_1=0$, $\lambda_{2,3}=(-1\pm \sqrt{5})/2$).

For two subspaces $|z| \geq 1$, there also exist two lines of equilibria. The first one ($x_1=0$, $y_1=0$, $|z|\geq 1$) represent a line of saddle-like equilibria (the green dashed lines in Fig.~\ref{fig3}~(a)) with eigenvalues $\lambda_1=0$, $\lambda_{2,3}=(-1\pm \sqrt{5})/2$. The second line of equilibria (the red solid lines in Fig.~\ref{fig3}~(a)) is an attractor and consists of steady states ($x_2=1$, $y_2=0$, $|z| \geq 1$) with the corresponding eigenvalues $\lambda_1=0$, $\lambda_{2,3}=(-1\pm i\sqrt{3})/2$. 

In conclusion, changing $z$ and keeping parameter values to be fixed one implements the bifurcation transition at $z=\pm1$ when the central segment of the line of equilibria ($x_1=0$, $y_1=0$) changes the stability at the same moment with the coexisting lines of equilibria ($x_2=\beta$, $y_2=0$). If system (\ref{transcritical_model}) involves tanh-function (\ref{beta_tanh}), the described transition is transformed into the transcritical bifurcation without parameters [Fig.~\ref{fig3}~(b)]: two lines of equilibria ($x_1=0$, $y_1=0$, $z\in(-\infty; \infty)$) and ($x_2=\beta$, $y_2=0$,$z\in(-\infty; \infty)$) intersect at $z=\pm1$ and exchange the stability. 

For non-zero values of the parameter $\delta$ the lines of equilibria disappear [Fig.~\ref{fig3}~(c)]. In such a case, the steady state $x=y=z=0$ becomes a single attractor in the phase space and the transcritical bifurcation can be induced only by varying the system parameter $\beta_1$. However, for small values $\delta$ the phase trajectories describing motion to the stable fixed point can trace the attracting segments of the lines of equilibria initially existed at $\delta=0$ (see the trajectories in Fig.~\ref{fig3}~(c)). 

\section{Saddle-node bifurcation}
Suppose that the switch in Fig.~\ref{fig1}~(b) operates in state 2 while the nonlinear transformation is quadratic. Then the expression for the feedback voltage becomes $V_{\text{f}}=(1+Z+R_1W(\varphi))+V+mV^2$, where the state variable varies according the equation $d\varphi /dt =RI-k\varphi$ and $m=-1$. Then the dimensionless model of the circuit takes the form:
\begin{equation}
\left\lbrace
\begin{array}{l}
\dfrac{dx}{dt}=\alpha y,\\
\\
\dfrac{1}{\gamma}\dfrac{dy}{dt}=-y+\beta(z)-x^2,\\
\\
\dfrac{dz}{dt} = y-\delta z, \\
\end{array}
\right.
\label{saddle_node_model}
\end{equation}
where the function $\beta(z)$ has two options: Exp. (\ref{beta_piecewise}) or Exp. (\ref{beta_tanh}). System (\ref{saddle_node_model}) is studied for fixed parameters: $\alpha=\gamma=1$, $\beta_1=-1$, $\beta_2=1$, $k=5$. 

\begin{figure}
\centering
\includegraphics[width=0.5\textwidth]{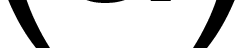} 
\caption{Saddle-node bifurcation without parameters in the phase space of system (\ref{saddle_node_model}) for memristive nonlinearities (\ref{beta_piecewise}) (panel (a)) and (\ref{beta_tanh}) (panel (b)) and the absence of the forgetting effect, $\delta=0$. The attracting segments of the lines of equilibria are marked by solid red lines while the repelling ones are marked by green dashed lines. Phase trajectories are black arrowed curves. The right insets schematically show the manifolds of equilibria as a projection on the plane $(x,z)$; (c)  Phase trajectory in system  (\ref{saddle_node_model})-(\ref{beta_tanh}) for $\delta=10^{-3}$ ending at infinity. Other system parameters are: $\alpha=\gamma=1$, $\beta_1=-1$,$\beta_2=1$, $k=5$.}
\label{fig4}
\end{figure}

First, system (\ref{saddle_node_model}) is considered for piecewise-smooth function $\beta(z)$ (function (\ref{beta_piecewise})) and $\delta=0$. Then in case $|z| < 1$ the system has no equilibria while two lines of equilibria ($x_{1,2}=\pm\sqrt{\beta_2}$, $y_{1,2}=0$, $|z| \geq 1$) exist for two subspaces $|z| \geq 1$ [Fig.~\ref{fig4}~(a)]. The first line of equilibria ($x_1=1$, $y_1=0$, $|z| \geq 1$) characterised by the eigenvalues $\lambda_1=0$, $\lambda_{2,3}=(-1\pm i\sqrt{7})/2$ is the system attractor. The second line of equilibria ($x_1=-1$, $y_1=0$, $|z| \geq 1$) exhibits the saddle-equilibrium properties (the corresponding eigenvalues are $\lambda_1=0$, $\lambda_{2,3}=(-1\pm 3)/2$).

In summary, changing $z$ and keeping parameter values to be fixed one implements the bifurcation transition at $z=\pm1$ when two equilibria ($x=\pm 1$, $y=0$) appear: stable and unstable ones. If system (\ref{saddle_node_model}) involves tanh-function (\ref{beta_tanh}), the described transition is transformed into the saddle-node bifurcation without parameters [Fig.~\ref{fig4}~(b)]: two lines of equilibria ($x=\pm1$, $y=0$, $|z|\geq 1$) appear at $z=\pm1$. The first line of equilibria is an attractor while the second one is unstable. 

For non-zero values of the parameter $\delta$ the lines of equilibria disappear. In such a case, the system has no equilibria. As demonstrated in Fig.~\ref{fig4}~(c), the phase trajectory can trace the attracting lines of equilibria initially existed at $\delta=0$, but after that the trajectories come to infinity. 

\section*{Conclusions}
\label{conclusions}
The circuit model proposed in the current paper represents a universal dynamical system for the implementation of bifurcations without parameters related to basic bifurcations of steady states: the pitchfork, transcritical and saddle-node bifurcations. The demonstrated bifurcations are associated with the existence of lines of equilibria. The transcritical and saddle-node bifurcations imply the coexistence of a stable and unstable line of equilibria. The pitchfork bifurcation realises the transition to the regime of bistability which consists in the coexistence of two  lines of equilibria attracting the phase trajectories and an unstable line of equilibria between them. 

The impact of the memristor forgetting effect is manifested in the contraction of limit sets along the OZ-axis. Then the resulting dynamics is determined by the limit sets crossing the plane $z=0$ in the phase space. If any line of equilibria intersect the plane $z=0$, it transforms into an isolated fixed point with the same stability properties.  Resultantly, a continuous dependence of the oscillation characteristics on the initial condition $z_0$ disappears. A similar character of the forgetting effect action was reported in earlier publications\cite{korneev2020,korneev2021,korneev2021-2}.

The transcritical bifurcation without parameter was discussed before \cite{riaza2018}. However, we are not aware of any models of dynamical systems exhibiting the pitchfork and saddle-node bifurcations without parameters. Apparently, the current paper is the first publication where the occurrence of the pitchfork and saddle-node bifurcations without parameters is analysed.
\begin{acknowledgements}
V.V.S. and A.S.Z. acknowledge support by the Deutsche Forschungsgemeinschaft (DFG, German Research Foundation) -- Projektnummer -- 163436311-SFB-910.
\end{acknowledgements}


\begin{thebibliography}{10}
\providecommand{\url}[1]{{#1}}
\providecommand{\urlprefix}{URL }
\expandafter\ifx\csname urlstyle\endcsname\relax
  \providecommand{\doi}[1]{DOI~\discretionary{}{}{}#1}\else
  \providecommand{\doi}{DOI~\discretionary{}{}{}\begingroup
  \urlstyle{rm}\Url}\fi

\bibitem{ascoli2013-2}
Ascoli, A., Senger, V., Tetzlaff, R.: Memristor model comparison.
\newblock IEEE Circuits and Systems Magazine \textbf{13}(2), 89--105 (2013)

\bibitem{botta2011}
Botta, V., N{\'e}spoli, C., Messias, M.: Mathematical analysis of a third-order
  memristor-based Chua's oscillator.
\newblock TEMA Tend. Mat. Apl. Comput. \textbf{12}(2), 91--99 (2011)

\bibitem{chang2011}
Chang, T., Jo, S., Kim, K., Sheridan, P., Gaba, S., Lu, W.: Synaptic behaviors
  and modeling of a metal oxide memristive device.
\newblock Applied Physics A \textbf{102}(4), 857--863 (2011)

\bibitem{chen2013}
Chen, L., Li, C., Huang, T., Chen, Y., Wen, S., Qi, J.: A synapse memristor
  model with forgetting effect.
\newblock Phys. Lett. A \textbf{377}(45-48), 3260--3265 (2013)

\bibitem{chua1971}
Chua, L.: Memristor -- the missing circuit element.
\newblock IEEE Trans. on Circuit Theory \textbf{CT-18}(5), 507--519 (1971)

\bibitem{chua2011}
Chua, L.: Resistance switching memories are memristors.
\newblock Appl. Phys. A \textbf{102}(4), 765--783 (2011)

\bibitem{chua1976}
Chua, L., Kang, S.: Memristive devices and systems.
\newblock Proceedings of the IEEE \textbf{64}(2), 209--223 (1976)

\bibitem{corinto2020}
Corinto, F., Forti, M., Chua, L.: Nonlinear Circuits and Systems with
  Memristors.
\newblock Springer (2020)

\bibitem{fiedler2000-2}
Fiedler, B., Liebscher, S.: Hopf bifurcation from lines of equilibria without
  parameters: II. systems of viscous hyperbolic balance laws.
\newblock SIAM J. Math. Anal. \textbf{31}(6), 1396--1404 (2000)

\bibitem{fiedler2000-1}
Fiedler, B., Liebscher, S., Alexander, J.: Generic Hopf bifurcation from lines
  of equilibria without parameters: I. theory.
\newblock Journal of Differential Equations \textbf{167}(1), 16--35 (2000)

\bibitem{fiedler2000-3}
Fiedler, B., Liebscher, S., Alexander, J.: Generic Hopf bifurcation from lines
  of equilibria without parameters: III. binary oscillators.
\newblock International Journal of Bifurcation and Chaos \textbf{10}(7),
  1613--1621 (2000)

\bibitem{gotthans2015}
Gotthans, T., Petr{\v z}ela, J.: New class of chaotic systems with circular
  equilibrium.
\newblock Nonlinear Dynamics \textbf{81}(3), 1143--1149 (2015)

\bibitem{gotthans2016}
Gotthans, T., Sprott, J., Petrzela, J.: Simple chaotic flow with circle and
  square equilibrium.
\newblock International Journal of Bifurcation and Chaos \textbf{26}(8),
  1650,137 (2016)

\bibitem{guseinov2021}
Guseinov, D., Matyushkin, I., Chernyaev, N., Mikhailov, A., Pershin, Y.:
  Capacitive effects can make memristors chaotic.
\newblock Chaos, Solitons and Fractals \textbf{144}, 110,699 (2021)

\bibitem{jafari2016}
Jafari, S., Sprott, J., Pham, V.T., Volos, C., Li, C.: Simple chaotic 3d flows
  with surfaces of equilibria.
\newblock Nonlinear Dynamics \textbf{86}(2), 1349--1358 (2016)

\bibitem{korneev2017-2}
Korneev, I., Semenov, V.: Andronov--Hopf bifurcation with and without parameter
  in a cubic memristor oscillator with a line of equilibria.
\newblock Chaos \textbf{27}(8), 081,104 (2017)

\bibitem{korneev2021}
Korneev, I., Semenov, V., Slepnev, A., Vadivasova, T.: Complete synchronization
  of chaos in systems with nonlinear inertial coupling.
\newblock Chaos, Solitons and Fractals \textbf{142}, 110,459 (2021)

\bibitem{korneev2020}
Korneev, I., Slepnev, A., Vadivasova, T., Semenov, V.: Forced synchronization
  of an oscillator with a line of equilibria.
\newblock Eur. Phys. J. Special Topics \textbf{229}(12), 2215--2224 (2020)

\bibitem{korneev2021-2}
Korneev, I., Slepnev, A., Vadivasova, T., Semenov, V.: Subcritical
  Andronov--Hopf scenario for systems with a line of equilibria.
\newblock Chaos \textbf{31}(7), 073,102 (2021)

\bibitem{korneev2017}
Korneev, I., Vadivasova, T., Semenov, V.: Hard and soft excitation of
  oscillations in memristor-based oscillators with a line of equilibria.
\newblock Nonlinear Dynamics \textbf{89}(4), 2829--2843 (2017)

\bibitem{liebscher2015}
Liebscher, S.: Bifurcation without Parameters, \emph{Lectures Notes in
  Mathematics}, vol. 2117.
\newblock Springer International Publishing (2015)

\bibitem{linn2014}
Linn, E., Siemon, A., Waser, R., Menzel, S.: Applicability of well-established
  memristive models for simulations of resistive switching devices.
\newblock IEEE Transactions on Circuits and Systems I: Regular Papers
  \textbf{61}(8), 2402--2410 (2014)

\bibitem{messias2010}
Messias, M., Nespoli, C., Botta, V.: Hopf bifurcation from lines of equilibria
  without parameters in memristor oscillators.
\newblock International Journal of Bifurcation and Chaos \textbf{20}(2),
  437--450 (2010)

\bibitem{riaza2012}
Riaza, R.: Manifolds of equilibria and bifurcations without parameters in
  memristive circuits.
\newblock SIAM J. Appl. Math. \textbf{72}(3), 877--896 (2012)

\bibitem{riaza2016}
Riaza, R.: Transcritical bifurcation without parameters in memristive circuits.
\newblock SIAM J. Appl. Math. \textbf{78}(1), 395--417 (2018)

\bibitem{riaza2018}
Riaza, R.: Transcritical bifurcation without parameters in memristive circuits.
\newblock SIAM J. APPL. MATH. \textbf{78}(1), 395--417 (2018)

\bibitem{semenov2015}
Semenov, V., Korneev, I., Arinushkin, P., Strelkova, G., Vadivasova, T.,
  Anishchenko, V.: Numerical and experimental studies of attractors in
  memristor-based Chua's oscillator with a line of equilibria. Noise-induced
  effects.
\newblock Eur. Phys. J. Special Topics \textbf{224}(8), 1553--1561 (2015)

\bibitem{singh2019}
Singh, J., Raj, B.: An accurate and generic window function for nonlinear
  memristor models.
\newblock Journal of Computational Electronics volume \textbf{18}(2), 640--647
  (2019)

\bibitem{tetzlaff2014}
Tetzlaff, R. (ed.): Memristor and Memristive Systems.
\newblock Springer-Verlag New York (2014)

\bibitem{zhou2019}
Zhou, E., Fang, L., Yang, B.: A general method to describe forgetting effect of
  memristors.
\newblock Phys. Lett. A \textbf{383}(11), 942--948 (2019)

\end{thebibliography}

\end{document}